\documentclass[conference]{IEEEtran}
\IEEEoverridecommandlockouts
\usepackage{cite}
\usepackage{amsmath,amssymb,amsfonts}
\usepackage{algorithmic}
\usepackage{graphicx}
\usepackage{textcomp}
\usepackage{xcolor}
\usepackage{booktabs}
\usepackage{multirow}
\def\BibTeX{{\rm B\kern-.05em{\sc i\kern-.025em b}\kern-.08em
    T\kern-.1667em\lower.7ex\hbox{E}\kern-.125emX}}
\begin{document}

\title{Instrumental Text-to-Music Generation with Auxiliary Conditioning Branches}

\author{\IEEEauthorblockN{Junyoung Koh}
\IEEEauthorblockA{\textit{Department of Artificial Intelligence}, \textit{Yonsei University} \\
\textit{MAAP} \\
\textit{KRAFTON} \\
Seoul, Republic of Korea \\
solbon1212@yonsei.ac.kr}}

\maketitle

\begin{abstract}

Text-to-music generation has advanced rapidly, with modern autoregressive and diffusion-based models producing convincing music from natural-language prompts. However, much of this progress relies on large-scale training data and external pretraining, making it difficult to isolate which design choices remain effective when data and pretraining are controlled. We study this setting using a Diffusion Transformer backbone with lyric and timbre conditioning, adapted to an instrumental-only text-to-music task in which the auxiliary lyric and timbre branches receive only degenerate conditioning signals. Through controlled ablations, we find that models retrained without these branches score lower across AudioBox aesthetics, LLM-as-judge, and human MOS, and that reinvesting the saved parameters as additional DiT depth recovers only marginally. This suggests the auxiliary branches may act as training-time architectural anchors whose contribution goes beyond their explicit conditioning content. We validate the same model through comparisons with external instrumental baselines and through our submission to the ICME 2026 Academic Text-to-Music (ATTM) Grand Challenge, where our Performance submission ranked first under both the objective metrics and the subsequent organizer-administered MOS over 35 raters, attaining the highest overall MOS across all challenge submissions, while our Efficiency submission was a finalist that tied for second under the objective metrics.

\end{abstract}


\section{Introduction}
\label{sec:intro}

Generative modeling of music conditioned on text has advanced substantially, with autoregressive models such as MusicGen \cite{copet2024musicgen} and MusicLM \cite{agostinelli2023musiclm} and diffusion-based models such as AudioLDM2 \cite{liu2023audioldm2}, Stable Audio \cite{evans2024stableaudio}, and MeLoDy \cite{lam2024melody} now able to synthesize convincing instrumental audio from natural-language prompts. This progress is driven by stronger backbones such as the Diffusion Transformer \cite{peebles2023dit} with flow matching \cite{lipman2023flow}, and by richer LLM-generated captions \cite{doh2023lpmusiccaps,manco2025jamendo}. However, much of this progress relies on large-scale training data and external pretraining, making it difficult to isolate which design choices remain effective once data and pretraining are controlled.

We study this setting using a Diffusion Transformer backbone with lyric and timbre conditioning, specifically the ACE-STEP 1.5 architecture~\cite{acestep2025}, adapted to an instrumental-only text-to-music task. ACE-STEP 1.5 was developed for vocal song generation, where its lyric and timbre encoders consume real lyrics and a reference vocal or instrumental timbre. Whether such a vocal-centric backbone transfers cleanly to instrumental-only generation, and whether its auxiliary branches still play a useful role under degenerate inputs, is not obvious a priori.

Through controlled architectural ablations, we find that retraining the model without these branches consistently degrades perceptual quality measured by AudioBox aesthetics~\cite{tjandra2025audiobox}, an LLM-as-judge protocol, and single-rater human MOS, while validation MSE and CLAP-style scores are much less sensitive to the change. Inference-time zeroing of either branch is much milder, suggesting that the encoder contribution is realized primarily during training. Even when we reallocate the parameters saved by removing the branches into additional DiT depth, so that the resulting model matches the original parameter count, the gap only partially closes, suggesting that the auxiliary branches provide an architectural benefit beyond their explicit conditioning content.

We validate these findings through comparisons with external instrumental text-to-music baselines, namely Stable Audio Open and three sizes of MusicGen. We additionally submit our model to the ICME 2026 Academic Text-to-Music (ATTM) Grand Challenge~\cite{hsieh2026academic}, where two tracks, an Efficiency Track capped at 500M parameters and an unrestricted Performance Track, share the same training dataset, a 457-hour subset of MTG-Jamendo~\cite{bogdanov2019mtg}, and the same set of allowed pretrained components. Building on the ACE-STEP backbone, we combine Min-SNR-$\gamma$ loss weighting~\cite{hang2023minsnr}, adaptive timestep sampling, random segment cropping, and post-hoc EMA averaging~\cite{karras2024edm2} on the training side, and a guidance interval~\cite{kynkaanniemi2024guidanceinterval} with a tuned EMA window on the inference side; complementary training-time and inference-time ablations quantify each component's contribution. We apply this recipe per track, training a 499M DiT for the Efficiency Track and a 2.4B DiT for the Performance Track. Our Performance model ranked first under both the objective metrics and the subsequent organizer-administered MOS, and our Efficiency model was a finalist that tied for second under the objective metrics.

\section{Related Work}
\label{sec:related}

\textbf{Text-to-Music Generation.} Existing approaches fall into two categories. Autoregressive models such as MusicGen \cite{copet2024musicgen} and MusicLM \cite{agostinelli2023musiclm} generate discrete audio tokens, while diffusion-based models such as AudioLDM2 \cite{liu2023audioldm2}, Stable Audio \cite{evans2024stableaudio}, and MeLoDy \cite{lam2024melody} operate in continuous latent spaces. Recently, ACE-STEP \cite{acestep2025} has combined a Diffusion Transformer \cite{peebles2023dit} with flow matching \cite{lipman2023flow} in a compressed audio latent space, achieving strong results with a modular architecture.

\textbf{Audio Captioning for Generation.} Caption quality and style significantly impact text-conditioned generation. LLM-generated captions~\cite{koh2025jamendoqa,koh2026jamendomtqa} have emerged as an effective way to augment music datasets, first shown by LP-MusicCaps~\cite{doh2023lpmusiccaps} and later scaled by JamendoMaxCaps~\cite{manco2025jamendo}, which provides 362K Jamendo captions produced with Qwen2-Audio. The ATTM challenge follows the same trend by releasing its own official Qwen2-Audio captions, which ensures style consistency between training captions and test prompts.

\textbf{Diffusion Training Strategies.} Several techniques have been proposed to improve diffusion model training. Min-SNR-$\gamma$ weighting \cite{hang2023minsnr} addresses timestep imbalance by clamping loss weights based on signal-to-noise ratio. EDM2 \cite{karras2024edm2} introduces post-hoc EMA reconstruction for improved checkpoint selection without retraining. Guidance interval \cite{kynkaanniemi2024guidanceinterval} restricts classifier-free guidance to mid-range timesteps, improving both sample quality and distribution fidelity.

\section{Method}
\label{sec:system}

We adopt the ACE-STEP 1.5 architecture \cite{acestep2025} as our backbone, a strong open-source music DiT model whose modular design consists of an audio VAE, a text encoder, a small condition encoder, and a DiT decoder. We train two instances at different scales, one per track: a 499M Efficiency model that uses 99.8\% of the Efficiency Track's 500M parameter cap, and a 2.4B Performance model. Both models use the same training and inference strategies; full configurations are in Table~\ref{tab:models}.

\textbf{Frozen pretrained components.} We use a frozen AutoencoderOobleck \cite{evans2024stableaudio} as the audio VAE; it maps 48\,kHz stereo audio to 25\,Hz, 64-channel latents at $1920{:}1$ temporal compression. A frozen Qwen3-Embedding-0.6B \cite{zhang2025qwen3embedding} serves as the text encoder, producing 1024-dimensional embeddings.

\textbf{Trained components.} The ATTM challenge is instrumental-only, so there are no real lyric or reference-audio inputs available to ACE-STEP's lyric and timbre branches. We therefore feed the lyric branch a single placeholder token ``[Instrumental]'' and the timbre branch a zero tensor. The condition encoder is a small transformer stack that further processes three conditioning streams, namely text, lyric, and timbre, and packs them into a single sequence for cross-attention. The lyric branch uses 6 layers in the Efficiency model and 8 in the Performance model, and the timbre branch uses 3 and 4 layers respectively. The lyric and timbre sub-modules therefore see constant inputs at every step. The DiT decoder is the main generative model, with alternating full and sliding-window attention layers, Grouped Query Attention \cite{ainslie2023gqa}, Adaptive Layer Normalization \cite{peebles2023dit} for timestep conditioning, and patch-based input and output projections with stride-2 convolutions.

\begin{table}[htbp]
\caption{Model Configurations for Both Tracks}
\begin{center}
\begin{tabular}{lcc}
\toprule
& \textbf{Efficiency} & \textbf{Performance} \\
\midrule
Hidden / Intermediate & 1152 / 3072 & 2048 / 6144 \\
DiT layers & 16 & 24 \\
Attention heads (KV heads) & 9 (3) & 16 (8) \\
Head dimension & 128 & 128 \\
Lyric encoder layers & 6 & 8 \\
Timbre encoder layers & 3 & 4 \\
Sliding window size & 128 & 128 \\
\textbf{Total parameters} & \textbf{499.0M} & \textbf{2,419.1M} \\
\midrule
Batch size $\times$ accumulation & 16$\times$4 & 8$\times$8 \\
Effective batch size & 64 & 64 \\
Training epochs & 120 & 120 \\
Hardware & 1$\times$A100-80GB & 1$\times$A100-80GB \\
Approx. GPU-hours & $\sim$20h & $\sim$60h \\
\bottomrule
\end{tabular}
\label{tab:models}
\end{center}
\end{table}

\section{Training Strategies}
\label{sec:training}

\textbf{Min-SNR-$\gamma$ loss weighting.} Standard flow matching~\cite{lipman2023flow} weights every timestep equally, which over-emphasizes easy high-SNR steps and under-uses informative low-SNR ones~\cite{hang2023minsnr}. This imbalance is especially costly in our limited-data regime, where every informative timestep matters. We therefore apply Min-SNR-$\gamma$ weighting with $\gamma$ of $5.0$, $w(t) = \min(\text{SNR}(t), \gamma) / \text{SNR}(t)$ where $\text{SNR}(t) = ((1-t)/t)^2$, which clamps easy-step weights while keeping full weight on the hard ones.

\textbf{Adaptive timestep sampling.} Min-SNR-$\gamma$ reweights each \emph{drawn} timestep, but leaves the sampling distribution itself fixed. We additionally adapt the distribution to the model's current loss profile: we discretize $[0, 1]$ into 100 bins, track an EMA with $\beta = 0.99$ of per-bin losses, and sample bins proportional to their temperature-scaled loss EMA. Before enough statistics accumulate, we fall back to a logit-normal default with $\mu = -0.4$ and $\sigma = 1.0$. The two techniques are complementary: one rebalances each draw's weight, the other rebalances how often each timestep is drawn at all.

\textbf{Random segment cropping.} The training set is 30-second clips, but generation targets 10 seconds. We randomly crop a 10-second window of 250 latent frames at 25\,Hz at every step, which provides about $3\times$ augmentation and exposes the model to diverse temporal content per track across epochs.

\textbf{Post-hoc EMA checkpoint averaging.} We save model snapshots every 1{,}000 steps throughout training. After training, we identify a late-training window in which validation loss has stabilized and uniformly average all snapshots inside it, following EDM2~\cite{karras2024edm2}. For the Efficiency model we use steps 70K through 96K, which corresponds to roughly the last 30 epochs, since averaging only converged snapshots avoids dragging the model toward early under-trained weights.

\textbf{Stability and optimization.} We clamp each per-sample MSE at $20.0$ before averaging, so a single outlier cannot dominate a batch update; outliers are still trained on but their gradient contribution is bounded. Optimization uses AdamW with learning rate $3\!\times\!10^{-4}$, weight decay $0.01$, and $\beta = (0.9, 0.95)$, applied with a 200-step linear warmup followed by cosine annealing. Both models are trained for 120 epochs with gradient checkpointing, and we use a $15\%$ classifier-free guidance dropout rate.

\section{Inference Strategies}
\label{sec:inference}

\textbf{ODE sampling with a guidance interval.} We generate with 100-step Euler ODE sampling and apply classifier-free guidance, abbreviated CFG~\cite{ho2022cfg}, only inside a \textit{guidance interval}~\cite{kynkaanniemi2024guidanceinterval} $t \in [0.1, 0.9]$.

\textbf{CFG scale strategy.} The CFG scale produces a well-known trade-off between FAD and CLAP: higher scales sharpen the conditional distribution and improve prompt adherence but can over-constrain the output. Instead of committing to a single CFG scale, we sweep CFG on a held-out development set and select two operating points per track that bracket the trade-off curve.

\section{Data and Caption Analysis}
\label{sec:data}

\subsection{Training Data}
We use exclusively the MTG-Jamendo \texttt{raw\_30s} subset with organizer-provided vocal separation, which contains approximately 55,700 tracks for a total of 457 hours, as released for the challenge in \cite{hsieh2026academic}. Audio is preprocessed into VAE latents and text embeddings are cached as \texttt{.pt} files. A 2\% random split is held out for validation.

\subsection{Caption Strategy Comparison}
Caption quality proved to be the most impactful factor in our experiments. We evaluated three caption strategies, summarized in Table~\ref{tab:captions}.

\begin{table}[htbp]
\caption{Impact of caption strategy on val loss (Efficiency 414M, 60 epochs). $\Delta$ is val loss relative to the template baseline.}
\begin{center}
\begin{tabular}{p{3.8cm}cc}
\toprule
\textbf{Caption Source} & \textbf{Val Loss} & \textbf{$\Delta$} \\
\midrule
Template (tag-to-text rules) & 0.968 & -- \\
JamendoMaxCaps (42\% match) & 0.943 & $-$0.025 \\
Official Qwen2-Audio (100\%) & 0.943 & $-$0.025 \\
\bottomrule
\end{tabular}
\label{tab:captions}
\end{center}
\end{table}

Template captions are tag-derived sentences obtained by applying simple rules to the MTG-Jamendo tag set, for example ``an atmospheric rock track with guitar,'' whereas Qwen2-Audio captions are detailed paragraphs describing genre, tempo, mood, and instrumentation. Crucially, the challenge's test prompts are also generated by Qwen2-Audio, so using Qwen2-Audio training captions aligns the train and test text distributions. JamendoMaxCaps, also produced by Qwen2-Audio, reaches the same val loss but covers only $42\%$ of our training tracks, so we use the official organizer-provided captions for full coverage. The $-0.025$ val-loss improvement from switching to LLM-generated captions exceeds the gain from any single training technique we tested, underscoring the importance of aligning the training and test text distributions.

\section{Experimental Results and Discussion}
\label{sec:results}

\subsection{Training Convergence}
\begin{figure}[htbp]
\centerline{\includegraphics[width=\linewidth]{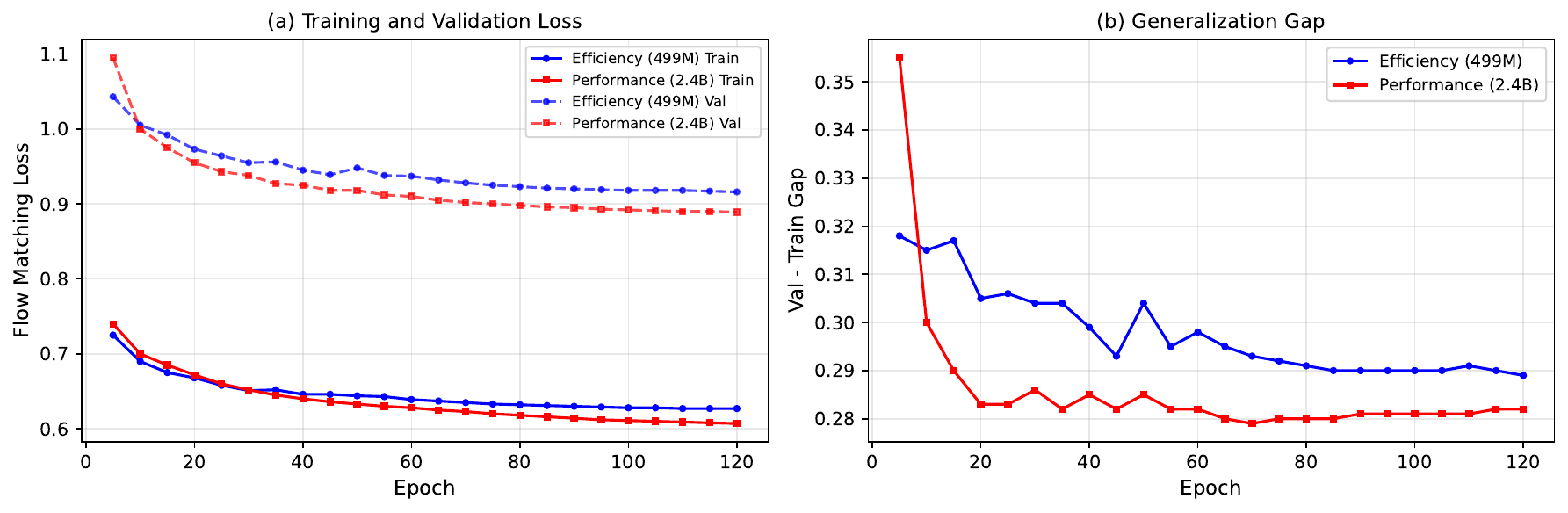}}
\caption{(a) Training and validation loss for both models over 120 epochs. (b) The generalization gap (val $-$ train) remains stable, indicating controlled overfitting.}
\label{fig:loss}
\end{figure}

Both models converge smoothly over 120 epochs, with a stable training and validation generalization gap of $0.29$ to $0.31$ that indicates no severe overfitting, as shown in Fig.~\ref{fig:loss}.

\subsection{Challenge Results}
\label{sec:quant_results}
Table~\ref{tab:results} reports the organizer's final-test evaluation on the 100 official prompts, with our two submitted entries (bold) listed alongside the top Efficiency-Track entries by MOS, the runner-up Performance entry, and the organizer-released MusicGen-small baseline.

\begin{table}[htbp]
\caption{Final-test results on the 100 official prompts. $\mathrm{MOS}_{\mathrm{all}}$ and $\mathrm{MOS}_{\mathrm{exp}}$ aggregate over all 35 raters and the 20-expert subset; Rank is by $\mathrm{MOS}_{\mathrm{all}}$ within each track. \textbf{Bold} rows are our entries.}
\begin{center}
\resizebox{\columnwidth}{!}{%
\begin{tabular}{lcccccc}
\toprule
\textbf{Entry} & \textbf{FAD}$\downarrow$ & \textbf{CLAP}$\uparrow$ & \textbf{CCS}$\uparrow$ & \textbf{MOS$_{\mathrm{all}}$}$\uparrow$ & \textbf{MOS$_{\mathrm{exp}}$}$\uparrow$ & \textbf{Rank} \\
\midrule
1st in Efficiency Track by MOS (e07)                  & 0.417 & 0.261 & 0.867 & 3.250 & 3.186 & 1 \\
2nd in Efficiency Track by MOS (e01)                  & 0.577 & 0.338 & 0.863 & 3.225 & 3.177 & 2 \\
3rd in Efficiency Track by MOS (e08)                  & --    & --    & --    & 3.119 & 3.044 & 3 \\
\textbf{Ours, Efficiency 499M (e05, CFG 8.0)}         & \textbf{0.487} & \textbf{0.305} & \textbf{0.800} & \textbf{2.969} & \textbf{2.929} & \textbf{4} \\
\textbf{Ours, Performance 2.4B (p05, CFG 7.0)}        & \textbf{0.514} & \textbf{0.306} & \textbf{0.800} & \textbf{3.344} & \textbf{3.327} & \textbf{1} \\
2nd in Performance Track by objective (p00)           & 0.557 & 0.311 & 0.796 & --    & --    & -- \\
MusicGen-small (organizer baseline)                   & --    & --    & --    & 3.538 & 3.425 & -- \\
\bottomrule
\end{tabular}}
\label{tab:results}
\end{center}
\end{table}

\textbf{Dry-run.} On the organizer's dry-run evaluation, our 414M Efficiency variants scored FAD~$0.51$--$0.56$ / CLAP~$0.31$--$0.32$ at CFG~$5.0$/$7.0$ and our 2.4B Performance scored FAD~$0.518$ / CLAP~$0.310$, against the best dry-run team at FAD~$0.404$ / CLAP~$0.323$; we then scaled the Efficiency model from 414M to 499M for the final submission to fully use the parameter budget.

\textbf{Final test.} The organizers selected CFG of 8.0 for our Efficiency entry and CFG of 7.0 for our Performance entry; the Efficiency choice sits comfortably on the CLAP plateau identified for that model in Fig.~\ref{fig:cfg_sweep}. Our 2.4B Performance entry achieves the top CCS in the Performance Track and the lowest FAD among Performance entries while CLAP stays within $0.005$ of the next model. Our 499M Efficiency entry ties for 2nd by objective metrics.

\textbf{Post-challenge MOS.} Under the organizer-administered MOS (Table~\ref{tab:results}, last two columns), our Performance submission attained the highest Overall score of any challenge entry in either track and won the Performance Track, while our Efficiency submission trailed the three Efficiency winners. The organizer-released MusicGen-small baseline outscored every submission, mirroring the training-data-scale gap also visible in our single-rater MOS (Sec.~\ref{sec:human_eval}).

\textbf{Self-eval.} Self-evaluation on the same 100 prompts with \texttt{msclap} CLAP and MERT-based FAD against an FMA-Medium reference we sample ourselves (absolute values not directly comparable to the organizer scale) gives Efficiency~499M FAD~$6.63$/$6.90$ and CLAP~$0.495$/$0.495$ at CFG~$7.0$/$8.0$, and Performance~2.4B FAD~$7.59$/$8.61$ and CLAP~$0.475$/$0.477$ at CFG~$7.0$/$10.0$.

\subsection{Incremental Strategy Ablation}
Table~\ref{tab:ablation} shows the cumulative impact of each strategy on the Efficiency track model.

\begin{table}[htbp]
\caption{Incremental Strategy Ablation (Efficiency Track). $\Delta$ is val loss relative to the baseline. Bold marks the lowest val loss reached.}
\begin{center}
\begin{tabular}{p{4.5cm}cc}
\toprule
\textbf{Configuration} & \textbf{Val Loss} & \textbf{$\Delta$} \\
\midrule
Baseline (template captions, self-processed audio, 60ep) & 0.968 & -- \\
\quad + Official vocal-separated audio & 0.960 & $-$0.008 \\
\quad + Official Qwen2-Audio captions & 0.943 & $-$0.025 \\
\quad + Post-hoc EMA averaging & 0.938 & $-$0.030 \\
\quad + Scale 414M $\rightarrow$ 499M & \textbf{0.916} & \textbf{$-$0.052} \\
\quad + Extended to 120 epochs & \textbf{0.916} & \textbf{$-$0.052} \\
\bottomrule
\end{tabular}
\label{tab:ablation}
\end{center}
\end{table}

Measured incrementally on top of the previous row, the largest single step is model scaling at $-0.022$, followed by caption quality at $-0.017$ incrementally or $-0.025$ versus the template baseline in Table~\ref{tab:captions}, official audio at $-0.008$, and EMA at $-0.005$ with zero additional training cost.

\subsection{Component-Wise Training Ablation}
\label{sec:training_ablation}

To isolate the contribution of each training-strategy component, we additionally train one variant per component on the Efficiency 499M model with exactly that component disabled while the others are held fixed, for up to 30 epochs.

\begin{table}[htbp]
\caption{Component-wise training ablation, step-aligned (Efficiency 499M). $\Delta$ is val loss relative to the baseline at the matched step; positive means the disabled component was helping.}
\begin{center}
\begin{tabular}{p{3.8cm}cc}
\toprule
\textbf{Disabled component} & \textbf{Val Loss} & \textbf{$\Delta$ (vs.\ base)} \\
\midrule
CFG dropout 0.15 $\rightarrow$ 0.05 & 0.962 & $+$0.032 \\
CFG dropout 0.15 $\rightarrow$ 0.30 & 0.961 & $+$0.029 \\
Min-SNR-$\gamma$ ($\gamma{=}5 \rightarrow 0$) & 0.960 & $+$0.026 \\
Adaptive timestep sampling off & 0.951 & $+$0.020 \\
\midrule
Random crop off (fixed center) & 0.892 & $-$0.041 \\
\bottomrule
\end{tabular}
\label{tab:train_ablation}
\end{center}
\end{table}

All four cleanly comparable components produce a strictly worse model when disabled, confirming that each contributes positively. The chosen CFG dropout of $15\%$ is a sweet spot: both halving the rate to $5\%$, with $\Delta = +0.032$, and doubling it to $30\%$, with $\Delta = +0.029$, hurt by similar amounts. The random-crop row is for completeness only: val uses the same crop policy, so disabling it makes val deterministic (fixed first 250 frames) and intrinsically easier.

\subsection{Inference Ablations}
\label{sec:inf_ablations}

We re-evaluate our Efficiency 499M model under three controlled inference-time changes, summarized in Table~\ref{tab:inf_ablation}, using the same checkpoint and 100 final-test prompts each time. Removing the guidance interval and applying CFG over the full $[0,1]$ drops CLAP by $-0.031$, which is the largest single inference-time degradation we see and confirms that CFG at extreme noise levels actively hurts text alignment~\cite{kynkaanniemi2024guidanceinterval}. Halving the ODE step count from 100 to 50 actually yields a marginal $+0.003$ improvement, and 200 steps is statistically tied with 100, so 50 steps is a strict efficiency win. Finally, the post-hoc EMA window matters: the last 10 or 30 epochs are interchangeable, but widening to all 96 snapshots from the full run drops CLAP by $-0.051$, because early under-converged snapshots pull the averaged model away from its operating point.

\begin{table}[htbp]
\caption{Inference Ablations (Efficiency 499M, CFG 7.0, msclap, $n=100$). $\Delta$ is CLAP relative to the baseline.}
\begin{center}
\begin{tabular}{lcc}
\toprule
\textbf{Configuration} & \textbf{CLAP}$\uparrow$ & \textbf{$\Delta$} \\
\midrule
Baseline (steps=100, GI=[0.1,0.9]) & 0.4958 & -- \\
\midrule
\textit{Guidance Interval} & & \\
\quad CFG over full [0,1] (GI off) & 0.4652 & $-$0.031 \\
\midrule
\textit{ODE Sampling Steps} & & \\
\quad 50 steps  & 0.4986 & $+$0.003 \\
\quad 200 steps & 0.4951 & $-$0.001 \\
\midrule
\textit{Post-hoc EMA Window} & & \\
\quad Last 10 epochs (16 snapshots) & 0.4961 & $+$0.000 \\
\quad Last 60 epochs (all 96 snap.) & 0.4455 & $-$0.051 \\
\bottomrule
\end{tabular}
\label{tab:inf_ablation}
\end{center}
\end{table}

\subsection{CFG Sweep}
\label{sec:cfg_sweep_results}
\begin{figure}[htbp]
\centerline{\includegraphics[width=\linewidth]{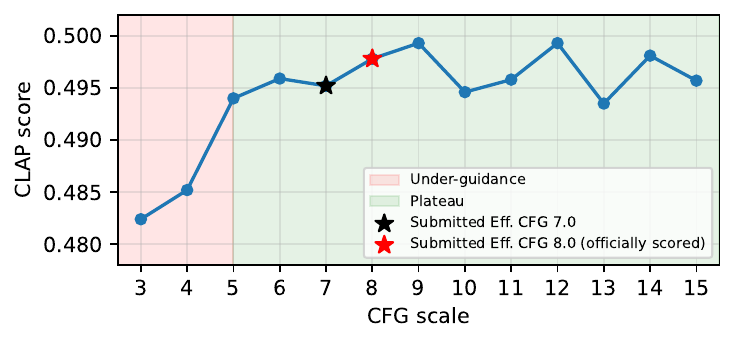}}
\caption{Sweep of CLAP against CFG on the Efficiency 499M model, evaluated with msclap on $n=100$ final-test prompts. CLAP rises monotonically up to CFG of about 5 and then plateaus within $\pm 0.003$. Stars indicate the two Efficiency-track variants we submitted, CFG 7.0 and 8.0; the organizers selected CFG 8.0 for the official scoring. Both sit on the CLAP plateau.}
\label{fig:cfg_sweep}
\end{figure}

To probe the relationship between CFG and CLAP at finer resolution, we sweep CFG in unit increments from 3 to 15 on the Efficiency 499M model and re-evaluate each with \texttt{msclap}, as plotted in Fig.~\ref{fig:cfg_sweep}. The curve shows a clear under-guidance regime up to CFG of about 5, where CLAP rises from $0.482$ to $0.494$, followed by a plateau above CFG~6 within $\pm 0.003$.

\begin{table*}[htbp]
\caption{Combined evaluation on 100 official final-test prompts (msclap CLAP, AudioBox, LLM-as-judge, single-rater human MOS; axes and protocol described in Sec.~\ref{sec:cond_ablation}--\ref{sec:human_eval}). \textbf{Bold} marks the best per column overall; \textcolor{blue}{\textbf{blue bold}} marks the best within the Efficiency block; \textcolor{red}{\textbf{red bold}} marks the best within the Performance block.}
\begin{center}
\setlength{\tabcolsep}{2.5pt}
\scriptsize
\begin{tabular}{lccc|cccc|ccc|ccc|ccc}
\toprule
& & & & \multicolumn{4}{c|}{\textbf{AudioBox}$\uparrow$} & \multicolumn{3}{c|}{\textbf{GPT-4o-audio}$\uparrow$} & \multicolumn{3}{c|}{\textbf{Gemini~3~Pro}$\uparrow$} & \multicolumn{3}{c}{\textbf{Human}$\uparrow$} \\
\textbf{Setup} & \textbf{Params} & \textbf{Data} & \textbf{msclap}$\uparrow$ & CE & CU & PC & PQ & CE & PQ & PA & CE & PQ & PA & Q & F & P \\
\midrule
\textit{Instrumental baselines} \\
Stable Audio Open~\cite{evans2024stableaudio}                  & 1.2B  & $\sim$7.3kh & 0.507 & 6.45 & 7.54 & 4.33 & 7.63 & 7.34 & 8.31 & 8.89 & 4.24 & 4.96 & 4.88 & 3.60 & \textbf{3.85} & 3.35 \\
MusicGen-small~\cite{copet2024musicgen}                        & 300M  & $\sim$20kh & \textbf{0.531} & 6.86 & 7.47 & 5.22 & 7.40 & 7.40 & 8.36 & 8.93 & 5.58 & 5.44 & 7.05 & 3.60 & 3.55 & 3.55 \\
MusicGen-medium~\cite{copet2024musicgen}                       & 1.5B  & $\sim$20kh & 0.528 & 6.85 & 7.53 & 5.25 & 7.44 & \textbf{7.42} & \textbf{8.42} & 8.92 & 5.85 & \textbf{5.72} & 7.20 & \textbf{3.85} & 3.75 & \textbf{3.80} \\
MusicGen-large~\cite{copet2024musicgen}                        & 3.3B  & $\sim$20kh & \textbf{0.531} & \textbf{7.12} & \textbf{7.69} & \textbf{5.42} & \textbf{7.58} & 7.38 & 8.40 & \textbf{8.95} & \textbf{5.95} & \textbf{5.72} & \textbf{7.43} & \textbf{3.85} & 3.75 & 3.55 \\
\midrule
\textit{Ours: Efficiency and conditioning ablations} \\
Efficiency Original, 16-layer DiT                              & 499M  & 457h & 0.4993 & \textcolor{blue}{\textbf{6.92}} & 7.40 & \textcolor{blue}{\textbf{5.17}} & \textcolor{blue}{\textbf{7.33}} & \textcolor{blue}{\textbf{7.41}} & \textcolor{blue}{\textbf{8.40}} & 8.89 & \textcolor{blue}{\textbf{3.08}} & 2.90 & \textcolor{blue}{\textbf{4.00}} & 2.95 & 3.10 & 2.75 \\
\quad without lyric encoder (inference)                        & 499M  & 457h & \textcolor{blue}{\textbf{0.5005}} & 6.66 & 7.39 & 4.81 & 7.26 & 7.35 & 8.38 & 8.88 & 3.07 & \textcolor{blue}{\textbf{3.12}} & 3.80 & \textcolor{blue}{\textbf{3.15}} & \textcolor{blue}{\textbf{3.30}} & \textcolor{blue}{\textbf{2.90}} \\
\quad without timbre encoder (inference)                       & 499M  & 457h & 0.4996 & 6.81 & \textcolor{blue}{\textbf{7.42}} & 4.96 & 7.33 & 7.39 & 8.37 & \textcolor{blue}{\textbf{8.95}} & 3.07 & 3.01 & 3.86 & 2.95 & 2.90 & 2.75 \\
\quad retrained without lyric \& timbre encoders, 16-layer DiT & 371M  & 457h & 0.4718 & 6.45 & 7.03 & 5.15 & 7.02 & 7.38 & 8.33 & 8.91 & 1.94 & 1.97 & 2.51 & 2.45 & 2.50 & 2.15 \\
\quad retrained without lyric \& timbre encoders, 23-layer DiT & 489M  & 457h & 0.4844 & 6.50 & 7.11 & 5.02 & 7.05 & 7.35 & 8.30 & 8.92 & 2.60 & 2.60 & 3.28 & 2.75 & 2.90 & 2.65 \\
\midrule
\textit{Ours: Performance and conditioning ablations} \\
Performance Original, 24-layer DiT                             & 2.4B  & 457h & 0.4737 & \textcolor{red}{\textbf{6.65}} & \textcolor{red}{\textbf{7.37}} & 4.89 & \textcolor{red}{\textbf{7.29}} & 7.37 & 8.35 & 8.87 & 3.34 & 3.43 & 4.25 & 2.95 & 3.20 & 2.85 \\
\quad without lyric encoder (inference)                        & 2.4B  & 457h & 0.4739 & 6.38 & 7.24 & 4.60 & 7.15 & 7.35 & 8.35 & 8.87 & 3.44 & 3.63 & \textcolor{red}{\textbf{4.36}} & 3.10 & \textcolor{red}{\textbf{3.40}} & 3.10 \\
\quad without timbre encoder (inference)                       & 2.4B  & 457h & \textcolor{red}{\textbf{0.4826}} & 6.53 & 7.31 & 4.70 & 7.25 & 7.37 & 8.37 & 8.90 & \textcolor{red}{\textbf{3.55}} & \textcolor{red}{\textbf{3.72}} & 4.32 & \textcolor{red}{\textbf{3.15}} & 3.25 & \textcolor{red}{\textbf{3.25}} \\
\quad retrained without lyric \& timbre encoders, 24-layer DiT & 1.8B  & 457h & 0.4776 & 6.47 & 7.06 & \textcolor{red}{\textbf{5.20}} & 7.01 & \textcolor{red}{\textbf{7.40}} & \textcolor{red}{\textbf{8.39}} & \textcolor{red}{\textbf{8.95}} & 2.03 & 2.00 & 2.81 & 2.25 & 2.55 & 2.25 \\
\bottomrule
\end{tabular}
\label{tab:cond_ablation}
\end{center}
\end{table*}

\subsection{Perceptual Evaluation and Architectural Conditioning Ablation}
\label{sec:cond_ablation}
We evaluate our two submitted models alongside four instrumental-only baselines, namely Stable Audio Open~\cite{evans2024stableaudio} and three sizes of MusicGen~\cite{copet2024musicgen}, on the same 100 final-test prompts. The evaluation uses three automatic measures: msclap CLAP, AudioBox aesthetics~\cite{tjandra2025audiobox}, and an LLM-as-judge protocol; we also report human MOS in Sec.~\ref{sec:human_eval}. The AudioBox axes report content enjoyment, content usefulness, production complexity, and production quality, abbreviated CE, CU, PC, and PQ. The LLM-as-judge protocol prompts two multimodal audio models, GPT-4o-audio at version \texttt{gpt-4o-audio-preview-2025-06-03} and Gemini~3~Pro at version \texttt{gemini-3-pro-preview}, to rate each 10-second clip on a 1 to 10 scale for content enjoyment, production quality, and prompt adherence, abbreviated CE, PQ, and PA.

We additionally probe the lyric and timbre encoders, which receive constant inputs (lyric ``[Instrumental]'', timbre zero tensor) yet still emit non-zero features into cross-attention. To distinguish a meaningful learned representation from a structural anchor the DiT co-adapts to, we run three controlled tests: \emph{(i)} inference-time zeroing of one encoder, \emph{(ii)} a from-scratch retrain with both encoders removed ($\sim$26\% fewer parameters), and \emph{(iii)} a capacity-matched retrain reallocating that budget into a 23-layer DiT (489M, still under the Efficiency cap).

The LLM-as-judge results in Table~\ref{tab:cond_ablation} highlight that the choice of judge model matters as much as the choice of metric. GPT-4o-audio produces a much narrower score range across models, suggesting limited discriminative power under this prompt. Gemini~3~Pro is far more discriminative: MusicGen leads on every axis, followed by SAO, with our submitted models and the inference-time conditioning ablations clustering below. Within each of our model groups, Gemini cleanly singles out the text-only architectures as the clearest losers. The AudioBox metrics also reflect this signal, but more weakly.

\textbf{Discussion.} Taken together with the mild inference-time zeroing effect, this pattern suggests the encoder contribution is realized primarily during training: the lyric and timbre branches play a structural role---either by reshaping the cross-attention pattern or by providing stable conditioning anchors---that more DiT depth alone cannot replicate, since the capacity-matched retrain recovers only marginally despite matching val MSE.

\subsection{Human Evaluation}
\label{sec:human_eval}
To complement the automatic metrics and LLM judges, a single author rates 20 of the 100 final-test prompts across all 13 setups in Table~\ref{tab:cond_ablation} on a 1 to 5 scale for quality (Q), prompt fidelity (F), and overall preference (P). The Q/F/P columns of Table~\ref{tab:cond_ablation} report the per-setup means.

Three patterns emerge. The MusicGen family leads at preference $3.55$--$3.80$, followed by Stable Audio Open at $3.35$ and our models at $2.7$--$2.9$, mirroring the training-data-scale gap. The text-only retrains are the worst among our models at $2.15$--$2.65$, agreeing with the AudioBox CE drop and with Gemini~3~Pro singling out the same architectures, and supporting the architectural-ablation finding that the lyric and timbre branches contribute as training-time anchors. Inference-time zeroing of either branch does not hurt human ratings and slightly improves them in several rows, consistent with the training-time effect interpretation.

\textbf{Discussion.} Across the conditioning ablation, the gap is visible on AudioBox aesthetics, Gemini~3~Pro, and human MOS but essentially invisible on validation MSE and msclap, and GPT-4o-audio would have masked it entirely. We therefore recommend triangulating across both metric families and judge models when comparing architecturally similar variants.

\subsection{Failure Case Analysis}
Three systematic failure modes show up under qualitative inspection. The first is \textit{instrument fidelity}, where distinctive instruments such as saxophone, harmonica, or vibraphone often render as generic timbres, reflecting their limited representation in the 457-hour set. The second is \textit{tempo precision}, where the model captures coarse energy levels but does not reliably hit specific BPM targets, since tempo is conveyed only implicitly through captions. The third is \textit{complex arrangements}, where prompts asking for four or more simultaneous instruments produce texturally muddy output and individual instruments are not separable. These primarily reflect the constrained data scale rather than architectural deficiencies.

\section{Limitations}
\label{sec:discussion}
Our model is trained from scratch on only 457 hours of audio under strict challenge constraints, our analyses use the official 100-prompt test set, and the human MOS used inside Table~\ref{tab:cond_ablation} for the conditioning ablation is single-rater on a 20-prompt subset; the organizer-released multi-rater MOS reported in Sec.~\ref{sec:human_eval} covers only the two submitted models and the organizer baseline, not the conditioning ablations. The structural role we attribute to the lyric and timbre sub-modules is observational; the precise mechanism is left to future work.

\section{Conclusion}

We studied a Diffusion Transformer text-to-music model with lyric and timbre conditioning, adapted to an instrumental-only task in which both auxiliary branches receive only degenerate inputs. Across AudioBox aesthetics, LLM-as-judge, and human MOS, retraining the model without these branches consistently degraded perceptual quality, while a capacity-matched deeper DiT recovered only marginally despite matching validation MSE. This suggests the auxiliary branches may act as training-time conditioning anchors that more DiT depth alone cannot easily replicate. We validated the same model through comparisons with external instrumental baselines and through our submission to the ICME 2026 ATTM Grand Challenge, where our Performance submission ranked \textbf{first under both the objective metrics and the subsequent organizer-administered MOS}, attaining the highest overall MOS across all challenge submissions, while our Efficiency submission was a finalist that tied for second under the objective metrics. These observations motivate further study of the architectural role of auxiliary conditioning branches in text-to-audio generation, including whether the same training-time anchor effect generalizes to other multi-modal architectures.

\bibliographystyle{IEEEbib}
\bibliography{icme2026references}

\end{document}